# Strong electromagnetic pulses generated in high-intensity laser-matter interactions

P Rączka[1*], J-L Dubois[2], S Hulin[2], M Rosiński[1], A Zaraś-Szydłowska[1] and J Badziak[1]

[1]Institute of Plasma Physics and Laser Microfusion, 01-497 Warsaw, Poland
[2]CELIA, University of Bordeaux-CNRS-CEA, 33405 Talence cédex, France

[*]Corresponding author: piotr.raczka@ifpilm.pl

**Abstract**. Results are reported of an experiment performed at the Eclipse laser facility in CELIA, Bordeaux, on the generation of strong electromagnetic pulses. Measurements were performed of the target neutralization current, the total target charge and the tangential component of the magnetic field for the laser energies ranging from 45 mJ to 92 mJ with the pulse duration approximately 40 fs, and for the pulse durations ranging from 39 fs to 1000 fs, with the laser energy approximately 90 mJ. It was found that the values obtained for thick (mm scale) Cu targets are visibly higher than values reported in previous experiments, which is argued to be a manifestation of a strong dependence of the target electric polarization process on the laser contrast and hence on the amount of preplasma. It was also found that values obtained for thin (µm scale) Al foils were visibly higher than values for thick Cu targets, especially for pulse durations longer than 100 fs. The correlations between the total target charge versus the maximum value of the target neutralization current, and the maximum value of the tangential component of the magnetic field versus the total target charge were analysed. They were found to be in very good agreement with correlations seen in data from previous experiments, which provides a good consistency check on our experimental procedures.

## 1. Introduction

The laser-target interaction at high power and high laser intensity may result in the emission of strong electromagnetic pulses (EMP) with frequencies in the range of tens of MHz to few GHz, as observed already in the seventies [1]. Such pulses may affect the data collection and disturb electronic devices used inside and outside of the experimental chamber, thus making it difficult to realise experiments in a safe and reliable way. With the advent of the petawatt power and MJ energy laser facilities a systematic effort was made to study the EMP phenomenon [2-6]. However, a full quantitative understanding has not been obtained and the problem continues to attract attention [7-14]. In particular, in a series of recent articles [7-9] authors presented a first detailed study of the electric polarization of the target and the resulting neutralization current as a source of EMP. They report on experiments at the fs Eclipse laser facility in CELIA, Bordeaux, in which they studied electric polarization of thick (mm scale) targets. It was found in particular that the charge generated on those targets and the maximum value of the neutralization current are directly proportional to the laser energy at fixed pulse duration and that these quantities are relatively insensitive to the pulse duration at fixed laser energy [7]. In [8] the results on the relation between the target charge and the maximum value of the EMP magnetic field were presented. It is of great interest to extend this analysis to the case of very thin (micrometer scale) foils, such as are commonly used in the laser-driven ion acceleration experiments. To this end a dedicated experiment was performed at the Eclipse facility by our group. Results of this experiment had been described in detail in [15]. In this note we summarize some of those results and compare them in some detail with the results of previous EMP experiments on Eclipse [7,8].

## 2. Experimental setup

The Eclipse laser at CELIA, Bordeaux, is a Ti:sapphire laser delivering a p-polarized beam with 807 nm central wavelength, focused by an f/5 off-axis parabolic mirror to the focal spot with the FWHM diameter of 10.50 μm. The laser beam was incident on the target along the target normal. The pulse duration was varied in the range 39 fs – 1000 fs, and the laser pulse energy was varied in the range 45 mJ – 95 mJ on target. The ns pedestal due to the amplified spontaneous emission (ASE) was found to be of 5 ns duration. The laser contrast for the majority of shots discussed in this note was $2\times10^{-6}$ at 150 ps before the peak, except for $8\times10^{-6}$ for the shots at pure Al foils and only $7\times10^{-5}$ for some of the initial shots at the thick Cu target. We used the Sophia experimental chamber.

Targets used in our experiment were designed so as to allow for a straightforward comparison between the EMP generated from thin and thick targets. A target holder in the form of a brass ring 14.0 mm in diameter, mounted on a thin brass wire stalk 28.0 mm long was used. Inside the brass ring the following targets were placed: (a) a massive Cu "pill" 10.1 mm in diameter and 1.0 mm thick; (b) an Al foil 6.0 μm thick, with 0.3 μm layer of polystyrene on the rear side, further denoted as "AlPs", pasted on the rear side of a Cu pill 10.1 mm in diameter and 1.0 mm thick, with 10 holes 1 mm in diameter for shots; (c) a pure Al 6.0 μm foil, placed between two Cu pills, each 0.50 mm thick, with 10 holes for shots. The target holder was mounted on a thin square Al plate (160.5 mm×160.5 mm, 4.2 mm thick) that was electrically grounded, to ensure the presence of a mirror charge and the dominance of the dipole component in the radiation pattern.

## 3. Neutralization current and the total target charge

In order to measure the target neutralization current the target holder stalk was directly connected via a coaxial cable to a high-bandwidth oscilloscope Teledyne LeCroy SDA 760Zi-A (6 GHz bandwidth, 20 GSa/s). The curves representing the temporal dependence of the target neutralization current were very similar for all three types of targets considered here, with a spike followed by few oscillations around the zero point and the timescale of ~1 ns. The net target charge was obtained by integrating over the current spike. The plot of the target charge as a function of the laser energy on target, for the pulse duration of approximately 40 fs, is shown in Fig. 1, together with the data obtained in previous experiments on Eclipse [7,8]. For parameters where three or more consistent data points were obtained in this experiment averages are shown with an error estimate, otherwise we just show a scatter plot.

We see that the values of the target charge obtained for the 1 mm Cu target are visibly higher than the values reported in [7] and [8] for 3 mm Cu target. For example, in [7] the value of (12±1) nC was obtained at 88 mJ and 50 fs, whereas we find (27±2) nC for 85 mJ and 39 fs and even (35±2) nC for 94 mJ and 55 fs. This is puzzling at first glance because both targets are thick enough to stop nearly all fast electrons generated on the front side of the target and in principle should give the same results for the target charge. The only other difference between these measurements is the laser contrast. Indeed, in the previous EMP campaigns on Eclipse the duration of the ASE pedestal was 2 ns and the contrast was $1\times10^{-7}$, whereas in our experiment the ASE pedestal was 5 ns and the contrast for the data points indicated in Fig. 1 by black dots was $2\times10^{-6}$. Thus it is our hypothesis that the difference in the target charge is a manifestation of a strong sensitivity of the target charge to the laser contrast and hence to the scale of the preplasma. This conjecture is supported by the fact that in the few shots made at 94 mJ and 55 fs that gave the extremely high value of the target charge the laser contrast was only $7\times10^{-5}$. Apart from this data point all our data for the 1 mm Cu target is in very good agreement with the direct proportionality of the target charge to the pulse energy, similarly to what was observed in [7] for the 3 mm Cu target, except that the slope is higher.

Turning to thin foil targets we see that the charge they generate is visibly higher than that generated from the 1 mm Cu target. For example, for the AlPs target we measure (32±2) nC at 86 mJ and 41 fs, which is 20% higher than the charge obtained for the 1 mm Cu target at similar laser parameters. The charge measured for the pure Al target appears to be higher by an even larger percentage, 50% or more, although this might be partially due to worse laser contrast ($8\times10^{-6}$).

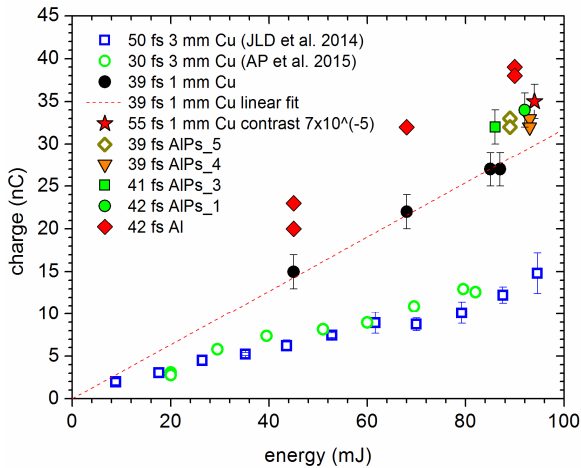

**Figure 1.** The total target charge as a function of the laser pulse energy on target, for the pulse duration close to 40 fs, as obtained in this experiment, compared with the data from previous experiments [7,8].

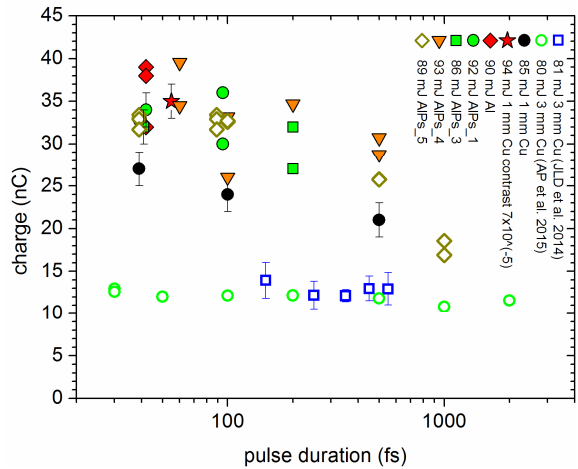

**Figure 2.** The target charge as a function of the pulse duration, for the pulse energy close to 90 mJ, as obtained in this experiment, compared with the data from previous experiments [7,8].

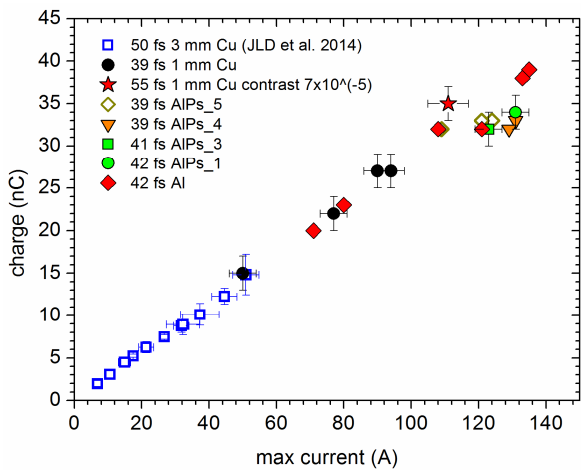

**Figure 3.** The correlation plot between the total target charge and the maximum value of the target neutralization current, for the data taken in the laser energy scan, for the pulse duration close to 40 fs, as obtained in this experiment, compared with the data from previous experiments [7,8].

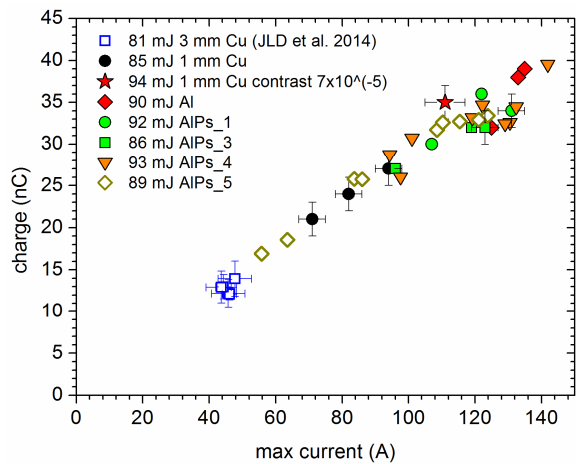

**Figure 4.** The correlation plot between the total target charge and the maximum value of the target neutralization current, for the data taken in the pulse duration scan, for the pulse energy close to 90 mJ, as obtained in this experiment, compared with the data from previous experiments [7,8].

In Fig. 2 we show the total target charge as a function of the pulse duration, for the pulse energy close to 90 mJ, together with the data obtained in previous experiments on Eclipse [7,8]. The following remarks can be made about the data from our experiment shown there: (1) for all values of the pulse duration the data from 1 mm Cu targets are visibly higher that the values reported in earlier experiments [7,8] (our hypothesis is that this is due to the difference in the laser contrast, as explained above); (2) the data for 1 mm Cu target show substantial dependence on the laser pulse duration, whereas data from earlier experiments are practically insensitive to the pulse duration (again, we interpret this qualitative difference as an interesting manifestation of the sensitivity to the laser contrast and the presence of a preplasma); (3) similarly as in the laser energy scan the charge obtained with thin targets is visibly higher that the charge generated with the thick targets.

As was mentioned above, the temporal profile of the neutralization current was nearly identical in all measurements, except for the maximum value of the current [15]. In such a case there should be a strong correlation between the values of the target charge and the maximum values of the return current. In Fig. 3 we show such correlation for the data from the pulse energy scan, and in Fig. 4 we display data obtained in the pulse duration scan (plots showing the data for the maximum values of the return current in these scans had been given in [15]). On both figures we see that the correlation is indeed very strong. This fact provides an important consistency check for our analysis. Furthermore, we see that the trend in the data obtained in [7] is in excellent agreement with the trend in the data from the present experiment, which provides a very strong argument that the difference between the experiments discussed above is not a result of some error but indeed a consequence of different physical conditions under which the measurements were made (i.e. the laser contrast). Finally, we note that several points in Fig. 3 and Fig. 4 corresponding to shots at the AlPs target lie below the line traced by other data points, which may indicate that there may be additional effects in the target charge generation and redistribution processes in such targets related to the presence of a Ps layer at the rear of the target.

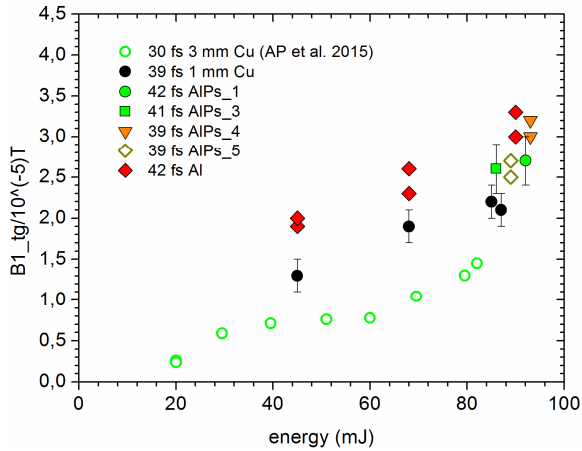 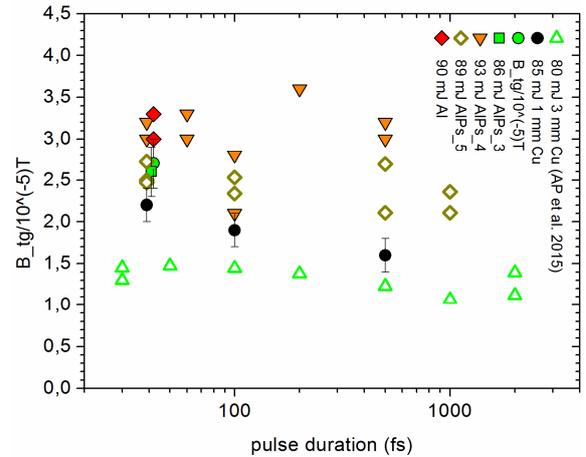

**Figure 5.** The maximum value of the tangential component of $B$ as a function of the laser energy, for the pulse duration close to 40 fs, as obtained in this experiment, compared with the data from previous experiment [8].

**Figure 6.** The maximum value of the tangential component of $B$ as a function of the pulse duration, for the laser energy close to 90 mJ, as obtained in this experiment, compared with the data from previous experiments [8].

## 4. Measurement of the EMP magnetic field

In order to characterize the EMP signal generated in laser-target interaction the magnetic field induction was measured inside the experimental chamber using a radiation-hardened Prodyn RB-230 B-dot probe, connected to a Prodyn BIB-100G balun. The Teledyne LeCroy oscilloscope connected via a 2 m long coaxial cable was used to collect the data. The probe was placed in a point located 214 mm behind the target, 41 mm above the target, and 55 mm to the right of the target, and set to measure the tangential component of the B field (relative to the natural cylindrical coordinate system in which the axis coincides with the vertical axis of the target holder). The data on $dB_{tg}/dt$ was processed by applying to the discrete

Fourier transform a band pass filter with 400 MHz lower cutoff and 7.5 GHz upper cutoff and correcting for the frequency-dependent attenuation of the cable, and then integrating to obtain $B_{tg}(t)$. The resulting signal has the form of a strong initial spike followed by decaying oscillations lasting approximately 200 ns, interspersed by what appears to be reflections of the original spike from the chamber walls and the conductive elements inside the experimental chamber. A detailed description of $B_{tg}(t)$ may be found in [15].

From the point of view of the EMP studies it is the initial spike that is most interesting, as it generally corresponds to the highest values of the field measured in the whole signal. Furthermore, it is also a signal originating directly from the laser-target interaction, undisturbed by any reflections or contamination by the chamber eigenmodes, and hence it is of more fundamental character. In order to characterize the magnitude of EMP a maximum value of the $B_{tg}$ was determined for each shot. In Fig. 5 we show the maximum value of $B_{th}$ as a function of the laser energy, with pulse duration kept close to 40 fs, and in Fig. 6 we show the same quantity as a function of the laser pulse duration, with pulse energy kept close to 90 mJ. On both figures the data obtained in [8] is also indicated for comparison. It should be noted that there is a normalization error in Fig. 4 of [8]: the values of $B_{tg}$ indicated in this figure have to be divided by a factor $2\pi$.

The following general statements can be made about the maximum values of $B_{tg}$ recorded in our experiment: (1) the values for the 1 mm Cu target are visibly higher than values reported in [8] for the 3 mm Cu target (after correcting it for the omitted factor of $2\pi$); (2) the values obtained for thin foil targets are systematically higher than values obtained for 1 mm Cu target. Interestingly, Fig. 5 looks quite similar to Fig. 1. In principle the same could be said about the Fig. 6 and Fig. 2, with an important exception of the measurements for thin targets at pulse durations longer than 100 fs, in which case we noticed a surprising effect that the magnitude of the initial spike retains approximately the same value with increasing pulse duration, despite a greatly reduced overall EMP signature. It should be noted that the maximum value of $B_{tg}$ measured for the 3 mm Cu target and longer pulses is also insensitive to the pulse duration, but apparently for a different reason, because in our measurements for 1 mm Cu target we see a clear evolution of $B_{tg}$ with the increasing pulse duration.

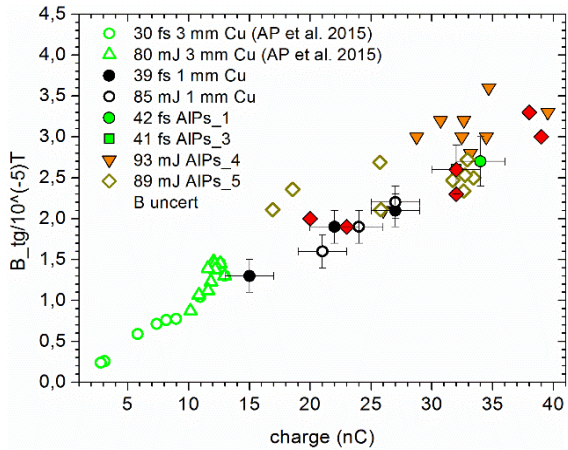 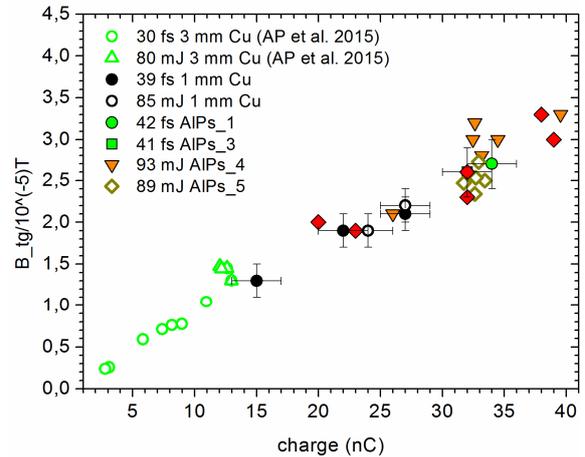

**Figure 7.** The correlation plot between the maximum value of the tangential component of the magnetic field and the total target charge, as obtained in this experiment in the laser energy scan at 40 fs pulse duration, and a pulse duration scan at 90 mJ laser energy, compared with the data reported in [8] for the 3 mm Cu target.

**Figure 8.** The same as in Fig. 7, excluding the data points corresponding to pulse duration longer than 100 fs.

The fact that Fig. 5 is quite similar to Fig. 1 suggests that it may be useful to study the correlation between the maximum value of $B_{tg}$ and the total target charge. Such a correlation is shown in Fig. 7, where all the data shown in previous figures for the laser energy scan and the pulse duration scan are included. It is interesting to compare this figure with Fig. 4 in [8]. We see that in general a relatively strong correlation is found. As we see, a nonlinear trend that seemed to be indicated by the data shown in [8] is not supported at higher values of the target charge. It is nevertheless evident that the scatter is much bigger than for example in Fig. 3 and Fig. 4 for the correlation between the charge and the maximum value of the current. To check whether this may be a consequence of an unusual behavior of the initial spike at larger values of the laser pulse we show in Fig. 8 the data from Fig. 7 with all points corresponding to the pulse duration longer than 100 fs removed from the plot. We see that the correlation in the data set restricted in this way is much stronger, which supports the hypothesis that for longer pulse durations other mechanisms of EMP generation might become relevant, apart from a straightforward neutralization of the target charge. Furthermore, the data from our experiment seems to be in excellent agreement with the data reported in [8], after correcting it for the omitted factor of $2\pi$. This provides yet another important cross check on the validity of our measurements and confirms high quality of our data analysis.

## 5. Conclusions

Summarizing, we performed measurements of the target neutralization current, the total target charge and the tangential component of the magnetic field of the resulting electromagnetic pulse for the laser energies ranging from 45 mJ to 92 mJ, with to pulse duration approximately 40 fs, and for the pulse durations ranging from 39 fs to 1000 fs, with the laser energy approximately 90 mJ. We found that values obtained for the 1 mm Cu target are visibly higher than values obtained in previous experiments [7,8] for the 3 mm Cu target, which we argue is a manifestation of a strong dependence on the laser contrast and hence on the amount of preplasma (the sensitivity of EMP to the presence of preplasma was noted in [16]). We also found that measurements for 6 μm Al foils give visibly higher results than for the thick Cu targets, which is especially true for pulse durations longer than 100 fs. Looking at plots of the target charge versus the maximum value of the neutralization current and the maximum value of the tangential component of the magnetic field versus the total target charge we find a very high degree of correlation, which provides an important check of the validity of our data collection and consistency of our data analysis. We also find very good agreement with the data obtained in previous experiments [7,8], after an error in normalization of the magnetic field reported in [8] is corrected.

**Acknowledgements**
The authors are grateful to V. Tikhonchuk for discussion and support and A. Poyé for comments on the results presented in [8]. This research is supported by the Polish National Science Centre grant Harmonia 2014/14/M/ST7/00024. Access to the Eclipse facility was made possible by the support obtained from LASERLAB-EUROPE, European Union's Horizon 2020 research and innovation programme, under the grant agreement no. 654148, project CNRS-CELIA002294. Our collaboration has profited greatly from the intellectual environment created by the COST Action MP1208.